# Multiparametric Cardiac $^{18}$F-FDG PET in Humans: Kinetic Model Selection and Identifiability Analysis

Yang Zuo, Ramsey D. Badawi, Cameron C. Foster, Thomas Smith, Javier E. López, Guobao Wang

*Abstract*—**Cardiac $^{18}$F-FDG PET has been used in clinics to assess myocardial glucose metabolism. Its ability for imaging myocardial glucose transport, however, has rarely been exploited in clinics. Using the dynamic FDG-PET scans of ten patients with coronary artery disease, we investigate in this paper appropriate dynamic scan and kinetic modeling protocols for efficient quantification of myocardial glucose transport. Three kinetic models and the effect of scan duration were evaluated by using statistical fit quality, assessing the impact on kinetic quantification, and analyzing the practical identifiability. The results show that the kinetic model selection depends on the scan duration. The reversible two-tissue model was needed for a one-hour dynamic scan. The irreversible two-tissue model was optimal for a scan duration of around 10-15 minutes. If the scan duration was shortened to 2-3 minutes, a one-tissue model was the most appropriate. For global quantification of myocardial glucose transport, we demonstrated that an early dynamic scan with a duration of 10-15 minutes and irreversible kinetic modeling was comparable to the full one-hour scan with reversible kinetic modeling. Myocardial glucose transport quantification provides an additional physiological parameter on top of the existing assessment of glucose metabolism and has the potential to enable single tracer multiparametric imaging in the myocardium.**

## I. Introduction

Positron emission tomography (PET) with the radiotracer $^{18}$F-fluorodeoxyglucose (FDG) is broadly used for imaging glucose metabolism [1, 2]. In clinical cardiology, FDG-PET is mainly applied to assess myocardial viability, myocardial inflammation, and other cardiac inflammatory diseases (e.g., cardiac sarcoidosis) [3-5]. Hibernating myocardium consists of viable muscles that are characterized by a perfusion-metabolism mismatch: low perfusion tracer uptake but high FDG uptake. In combination with PET perfusion imaging using a flow tracer such as $^{82}$Rb-chloride [6-10], $^{15}$O-water [11, 12], or $^{13}$N-ammonia [8, 13], FDG-PET assessment of myocardial viability has become a valuable clinical tool to identify this mismatch [14]. Patients with coronary artery disease who have hibernating myocardium usually benefit from additional surgical revascularization [15].

Standard cardiac FDG-PET uses static scanning and provides standardized uptake value (SUV) for the characterization of myocardial glucose metabolism. Alternatively, dynamic FDG-PET has also been investigated for quantitative myocardial imaging [16-18]. With compartmental modeling [19-22] or graphical analysis [23-27], most of these existing dynamic cardiac FDG-PET studies focused on quantitative evaluation of glucose metabolism. Nevertheless, the ability of dynamic FDG-PET for assessing blood-to-myocytes glucose transport has rarely been exploited for clinical applications in cardiac imaging.

We hypothesize that a glucose transport-metabolism mismatch exists in a hibernating myocardium, similar to the well-established flow-metabolism mismatch [3, 4]. The hypothesis is built upon the potential correlation between glucose transport and blood flow, which has been demonstrated for noncardiac tissues such as tumor [28-30] but not yet for the myocardium. Successful testing of this hypothesis may lead to a new and more accessible imaging solution using FDG-PET only to assess myocardial viability or inflammation without the need for a perfusion tracer. This FDG-only method has the potential to reduce imaging time, cost, and radiation exposure as compared to the traditional two-tracer methods (e.g., Rb-82 PET plus FDG-PET) in clinical use today [31, 32].

The main purpose of this study was to establish an effective and efficient dynamic imaging and kinetic analysis approach to quantifying myocardial glucose transport from dynamic FDG-PET in human patients. Standard dynamic FDG-PET scanning commonly lasts for one hour [33], and use both the irreversible and reversible two-tissue compartmental models for model selection with mixed results [22, 34]. In addition, the attention of these previous studies was mainly on the FDG net influx rate $K_i$ or the metabolic rate of glucose that characterize myocardial glucose metabolism [16-18, 22, 27], but not on glucose transport quantification.

In this paper, we revisited the kinetic model selection for dynamic cardiac FDG-PET imaging using cardiac patient scans.

This work was supported in part by National Institutes of Health (NIH) under grant no. R21 HL131385, American Heart Association under grant no. 15BGIA25780046. The work of J.E.L. is also supported in part by the Harold S. Geneen Charitable Trust Awards Program and the National Center for Advancing Translational Sciences, NIH, grant number UL1 TR001860 and linked award KL2 TR001859.

Y. Zuo, C.C. Foster, and G.B. Wang are with the Department of Radiology, University of California Davis Medical Center, Sacramento, CA 9817. (Email: yngzuo@ucdavis.edu, ccfoster@ucdavis.edu, gbwang@ucdavis.edu).

R.D. Badawi is with the Department of Radiology and Department of Biomedical Engineering, University of California Davis Medical Center, Sacramento, CA 9817. (Email: rdbadawi@ucdavis.edu).

T. Smith and J.E. López are with the Department of Internal Medicine, University of California Davis Medical Center, Sacramento, CA 9817. (Email: twrsmith@ucdavis.edu, drjlopez@ucdavis.edu).

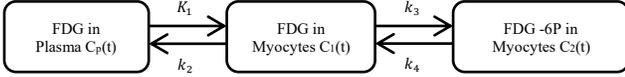

**Figure 1:** The reversible two-tissue compartment model for myocardial FDG kinetics. $K_1$ is the glucose transport rate from plasma to cardiac myocytes and $k_2$ is the transport rate from myocytes to plasma. FDG is phosphorylated by hexokinase in cells into FDG 6-phosphate with the rate $k_3$ and the process can be reversible by the rate $k_4$. The model becomes irreversible if $k_4=0$.

In particular, we investigated the effect of scan duration on model selection and its impact on quantification of myocardial kinetics with a focus on glucose transport quantification. A practical identifiability analysis was conducted to evaluate how reliable the myocardial kinetic parameters can be estimated. Our results provide supporting data and foundational protocols for the further testing of the myocardial glucose transport-metabolism mismatch in future studies using FDG-PET only for evaluating myocardial viability.

This paper is organized as follows. Section II describes the methods of dynamic FDG-PET data acquisition, kinetic modeling, model selection, and practical identifiability analysis for myocardial kinetic quantification. The results are then reported in Section III, followed by a detailed discussion of the findings and limitations of this study in Section IV. Finally, conclusions are drawn in Section V.

## II. METHODS

### A. Dynamic $^{18}$F-FDG PET/CT Data Acquisition

Fourteen patients who were scheduled for PET myocardial viability assessment consented into this study. The study was approved by the Institutional Review Board at the University of California, Davis. Each patient in the study first received glucose and/or insulin administration following a standard clinical protocol to ensure the highest quality of images. Patients then underwent a dynamic FDG-PET/CT scan followed by a standard static FDG-PET/CT scan on a GE Discovery ST PET/CT scanner in two-dimensional mode. The center of the scanner axial field of view was positioned at the heart of each patient. Patients received approximately 20 mCi $^{18}$F-FDG with bolus injection.

For the dynamic PET scan, list-mode data acquisition commenced immediately following the FDG injection. A low-dose transmission CT scan was performed before the dynamic PET scan for PET attenuation correction.

The raw PET data were binned into a total of 49 dynamic frames: 30 × 10s, 10 × 60s, and 9 × 300s (i.e., 60 minutes acquisition). Dynamic FDG-PET images were reconstructed using the standard ordered subsets expectation maximization (OSEM) algorithm with two iterations and 30 subsets as provided in the vendor software. Standard corrections for normalization, attenuation, dead-time, scatter, and random coincidences were applied. No motion correction was applied in this study.

### B. Kinetic Modeling of Dynamic Cardiac FDG-PET Data

*(1) Extraction of blood input functions and myocardial time activity curves (TACs).* Two 3D regions of interest (ROIs) were manually placed in the approximate centers the left ventricle (LV) and right ventricle (RV) to extract the image-derived blood input functions $C_{\text{LV}}(t)$ and $C_{\text{RV}}(t)$ in the unit of Bq/mL. An additional 17 ellipsoidal ROIs were placed within the 17 segments of myocardium according to the AHA-17 standard [18]. These segment ROIs were combined into a global myocardial ROI and used to extract a global myocardial TAC $C_{\text{T}}(t)$ using ROI mean. No postprocessing was applied to the TACs. Due to high noise of the dynamic data in individual segments, the analysis of this study was focused on global myocardial quantification.

*(2) Full Kinetic Modeling.* The reversible two-tissue (2T) compartmental model [35, 36] shown in Fig. 1 was first used to model the one-hour dynamic data. The corresponding ordinary differential equations of this 2T model are:

$$\frac{\mathrm{d}C_1(t)}{\mathrm{d}t} = K_1 C_{\text{LV}}(t) - (k_2 + k_3) C_1(t) + k_4 C_2(t), \quad (1)$$

$$\frac{\mathrm{d}C_2(t)}{\mathrm{d}t} = k_3 C_1(t) - k_4 C_2(t), \quad (2)$$

where $C_1(t)$ is the activity concentration of free FDG and $C_2(t)$ is the activity concentration of metabolized tracer in the myocardium tissue space. $K_1$ (mL/g/min) is the rate of glucose transport from the plasma to the tissue space; $k_2$ (/min) is the rate constant of tracer exiting the tissue space; $k_3$ (/min) is the rate constant of FDG being phosphorylated; $k_4$ (/min) is the rate constant of FDG-6P being dephosphorylated.

The total radioactivity that can be measured by PET is

$$C_{\text{T}}(t; \boldsymbol{\theta}) = (1 - v_{\text{LV}} - v_{\text{RV}})[C_1(t) + C_2(t)] \\ + v_{\text{LV}} C_{\text{LV}}(t) + v_{\text{RV}} C_{\text{RV}}(t), \quad (3)$$

where $v_{\text{LV}}$ and $v_{\text{RV}}$ denote the fractional blood volume parameters attributed from the LV and RV, respectively. $\boldsymbol{\theta} = [v_{\text{LV}}, v_{\text{RV}}, K_1, k_2, k_3, k_4]^T$ is a vector collecting all unknown parameters.

The model parameters are estimated by fitting a measured myocardial TAC $\{\check{C}_{\text{T}}(t_m)\}$ using the following nonlinear least-square optimization:

$$\hat{\boldsymbol{\theta}} = \arg \min_{\boldsymbol{\theta}} RSS(\boldsymbol{\theta}), \quad (4)$$

$$RSS(\boldsymbol{\theta}) = \sum_{m=1}^{M} w_m \left[\check{C}_{\text{T}}(t_m) - C_{\text{T}}(t_m; \boldsymbol{\theta})\right]^2, \quad (5)$$

where $t_m$ is the mid-point of the *m*th frame in a total of *M* time frames and $w_m$ is the weighting factor. In this work, we used a uniform weight (i.e. $w_m = 1$). $RSS(\boldsymbol{\theta})$ denotes the residual sum of squares of the curve fitting. The classic Levenberg-Marquardt algorithm with 100 iterations was used to solve the optimization problem in a way similar to our other work [37]. The fitting process was implemented using C/C++ programming. In this study, the initial value of $\boldsymbol{\theta}$ was set to $[0.1, 0.1, 0.1, 0.1, 0.1, 0.001]^T$. The lower bound was zero and the upper bound was $[1.0, 1.0, 2.0, 2.0, 1.0, 0.1]^T$.

For glucose transport quantification, the parameter of our interest is $K_1$. For glucose metabolism evaluation, the SUV at 4one-hour post-injection is a semi-quantitative measure. Hence, we also calculated the FDG net influx rate $K_i$ from the estimated micro parameters using the form of:

$$K_i = K_1 k_3 / (k_2 + k_3) \qquad (6)$$

to provide a quantitative measure of glucose metabolism.

*(3) Model selection.* For analyzing the dynamic cardiac FDG-PET data, we compared three different tracer kinetic models as listed in Table 1 : (1) 2T6P – the reversible two-tissue (2T) model with all kinetic parameters $[v_{LV}, \ v_{RV}, K_1, k_2, k_3, k_4]$ estimated, (2) 2T5P – the irreversible 2T model with $k_4 = 0$, i.e., the dephosphorylation process is neglected in the modeling, and (3) 1T4P - a simplified one-tissue (1T) model without modeling the phosphorylation and dephosphorylation processes. This 1T4P model is equivalent to the 2T model with $k_3 = 0$ and $k_4 = 0$. Different models were compared for statistical fit quality using the Akaike information criteria (AIC) [38, 39],

$$\text{AIC} = M \ln\left(\frac{RSS}{M}\right) + 2n + \frac{2n(n+1)}{M-n-1}, \qquad (7)$$

where $M$ denotes the number of time frames used in fitting and $n$ denotes the total number of unknown parameters, as listed in Table 1 for different models. Here AIC was corrected for finite sample sizes due to the ratio $\frac{M}{n} \leq 40$. A lower AIC value indicates a better selection of models [40].

*(4) Effect of scan duration on model selection and kinetic quantification.* In this study, we also investigated the effect of scan duration post tracer injection on kinetic model selection. The scan duration was varied from two minutes to one hour following the successive time frames used in the scanning protocol. For each scan duration, the three candidate models were compared for statistical fit quality evaluation using AIC. The impact on myocardial kinetic parameter quantification was evaluated as a function of scan duration. The kinetic parameters $K_1$ and $K_i$ estimated by 2T6P of the one-hour dynamic data were considered as the reference for the patient data.

### C. Practical Identifiability Analysis of Kinetic Quantification

We conducted a practical identifiability analysis using the Monte Carlo simulation approach to estimate the statistical properties associated with myocardial kinetic estimation, in a way similar to our previous work [41].

*(1) Determination of the TAC noise level.* The TAC noise level of the patient data in the myocardium is estimated using the noise model [42]:

$$\Delta C_T(t_m) = \frac{\check{C}_T(t_m) - C_T(t_m; \boldsymbol{\theta})}{\delta_m} \sim \text{Gaussian}(0, S_c) \qquad (8)$$

where the normalized residual activity $\Delta C_T(t_m)$ in the $m$th frame is calculated as the difference between the noisy measurement $\check{C}_T(t_m)$ and the fitted value $C_T(t_m; \boldsymbol{\theta})$. $\delta_m$ is the frame-dependent SD after normalization following the widely used model,

$$\delta_m = \sqrt{C_T(t_m; \boldsymbol{\theta}) \exp(\lambda t_m) / \Delta t_m} \qquad (9)$$

with $\lambda = \ln(2/T_{1/2})$ the decay factor, $T_{1/2}$ (min) the half-life of the radiotracer, and $\Delta t_m$ the duration of each frame. $S_c$ is the SD of the standardized Gaussian distribution. Its value was obtained by fitting the histogram of $\Delta C_T(t_m)$ using the standardized Gaussian distribution (see Fig. 5). $S_c \delta_m$ together represents the noise level in each time frame $m$.

**Table 1.** List of different kinetic models that are studied in this paper for dynamic cardiac FDG-PET kinetic modeling

| Model Type | Number of Unknown Parameters, $n$ | Kinetic Parameters to Be Estimated | Fixed Parameters |
|---|---|---|---|
| 2T6P | 6 | $v_{LV}, \ v_{RV}, K_1, k_2, k_3, k_4$ | N/A |
| 2T5P | 5 | $v_{LV}, \ v_{RV}, K_1, k_2, k_3$ | $k_4 = 0$ |
| 1T4P | 4 | $v_{LV}, \ v_{RV}, K_1, k_2$ | $k_3 = 0, k_4 = 0$ |

*(2) Computer simulation.* The estimated FDG kinetic parameter set $\boldsymbol{\theta}^0$ from the one-hour dynamic data with the 2T6P model was taken as the nominal set of parameters for each patient. Together with the patient's blood input function, the nominal parameter set was used to generate the noise-free myocardial TAC. Independently and identically distributed noise was then added to the noise-free TAC to generate $N = 1000$ realizations of noisy TACs. The noisy TACs were then fitted to obtain the noisy estimates of the kinetic parameters.

*(3) Evaluation metrics.* The normalized bias, standard deviation (SD), and root mean square error (RMSE) were calculated to evaluate the statistical properties of the kinetic parameter estimation,

$$\text{Bias}(\hat{\theta}_k) = \frac{\text{Mean}(\hat{\theta}_k) - \theta_k^0}{\theta_k^0}, \qquad (10)$$

$$SD(\hat{\theta}_k) = \frac{1}{\theta_k^0} \sqrt{\frac{1}{N-1} \sum_{n=1}^{N} [\hat{\theta}_k - \text{Mean}(\hat{\theta}_k)]^2}, \qquad (11)$$

$$MSE(\hat{\theta}_k) = \sqrt{\frac{1}{N} \sum_{n=1}^{N} \left(\frac{\hat{\theta}_k - \theta_k^0}{\theta_k^0}\right)^2}, \qquad (12)$$

where $\theta_k^0$ denotes the $k$th kinetic parameter in $\boldsymbol{\theta}^0$. $\text{Mean}(\cdot)$ represents the mean value of the noisy kinetic parameter estimates $\hat{\theta}_k$. The same simulation process was repeated for all the 10 patient data sets.

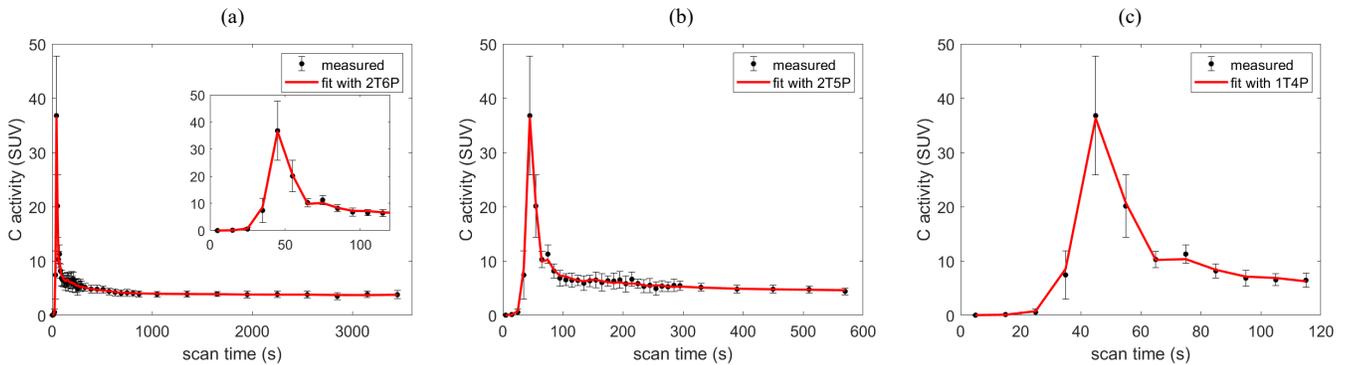

**Figure 2**. Example of global myocardial TAC fitting. **(a)** Fitting of a one-hour TAC with the 2T6P model; **(b)** Fitting of a 10min TAC using the 2T5P model; **(c)** Fitting of a 120s TAC using the 1T4P model.

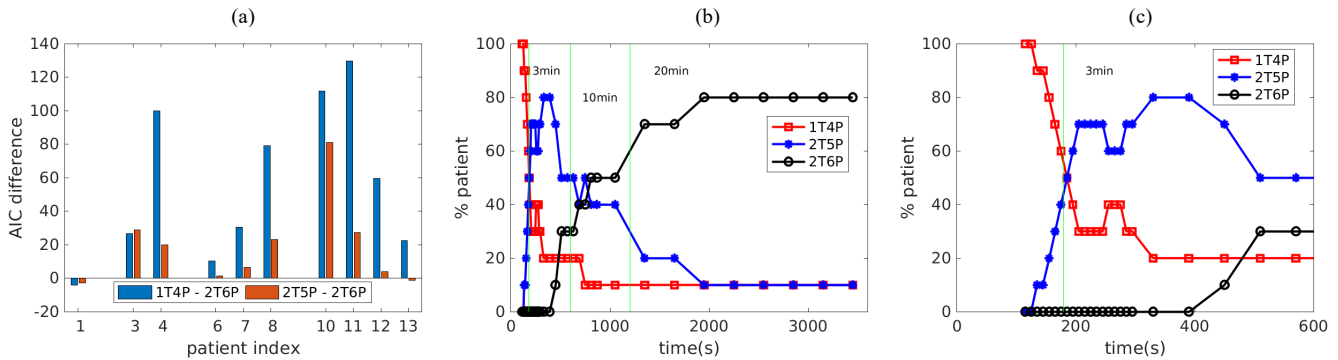

**Figure 3**. Comparison of different kinetic models using AIC. **(a)** AIC comparison in ten out of 14 patients that are applicable for this analysis. A positive AIC difference indicates that the 2T6P model provides a better TAC fitting quality than the comparison model; **(b-c)** Plots of the percentage of patients favoring a kinetic model over other models as a function of scan duration which varies from 2 minutes to one hour (b) and the zoom-in plots for the first 10 minutes (c).

## III. RESULTS

### A. Patient Characteristics

Ten out of 14 patients completed the one-hour scan. The remaining four patients had shorter scan duration, ranging from 30-45 minutes, because patients experienced discomfort. In this work, we only included the ten patients who underwent the complete one-hour dynamic scan for the data analysis. Of the ten patients, two were female (patient index #4 and #13 in figure 3) and three were diabetic (#1, #8, #13). Patients had age of 67±10 years in the range of 57-83 years, blood sugar level of 106±9 mg/dL in the range of 84-135 mg/dL, and body mass index of 30±5 kg/m² in the range of 22-39 kg/m².

### B. Examples of TAC Fitting

Figure 2 shows an example of measured global myocardial TAC of a patient with the error bars denoting the standard deviation of ROI activity quantification across the 17 segments in the myocardium. Figure 2(a) shows the fitting of the global myocardial TAC for the full one-hour data. The fitted TAC using the 2T6P model matched well with the noisy TAC. Figure 2(b) and 2(c) respectively show the fitting for the early-dynamic TAC of 10 minutes using the 2T5P model and the early-dynamic TAC of 120 seconds using the 1T4P model. All demonstrated good fitting in the respective time period.

### C. AIC Comparison of Kinetic Models for the One-hour Data

To evaluate the quantitative measure of statistical TAC fitting quality, the AIC differences between 2T6P and 2T5P or 1T4P are plotted in Fig. 3(a) for the one-hour data for all ten applicable patients. A positive AIC difference indicates that the 2T6P model provides a better TAC fitting quality than the 2T5P or 1T4P model, and vice versa for a negative value of AIC difference. In 8 out of 10 patients, the 2T6P model outperformed the 2T5P and 1T4P models. The 1T4P model and the 2T5P model were only favored in one patient. The comparison result demonstrated that 2T6P is an appropriate model in most of the patients for fitting the one-hour global TAC data.

### D. Effect of Scan Duration on Model Selection

Figure 3(b-c) shows the percentage of patients whose TAC fitting preferred the 2T6P, 2T5P, and 1T4P models as a function of scan duration. The scan duration varied from one hour to 2 minutes. In the range from 20 to 60 minutes, the 2T6P model was generally favored in 70-80% patients. As the scan duration was reduced to 3-15 minutes, the 2T6P model was decreasingly preferred and the 2T5P became more appropriate for fitting the TACs. When the scan duration was further reduced to 2-3 minutes, the 1T4P model was preferred for TAC fitting for most patients.

### E. Impact on Myocardial Kinetic Quantification

Figure 4(a) shows how the myocardial FDG $K_1$ values estimated by the 1T4P, 2T5P, and 2T6P models change as a function of the scan duration varying from 2 minutes to one hour in one example patient. Figure 4(b) summarizes the results from all the ten patients. Using $K_1$ of 2T6P from the full one-hour data as the reference, the $K_1$ estimation with the 2T6P model increasingly deviates as the scan duration is shortened. The difference in $K_1$ exceeded 5% once the scan duration was shorter than 30 minutes.

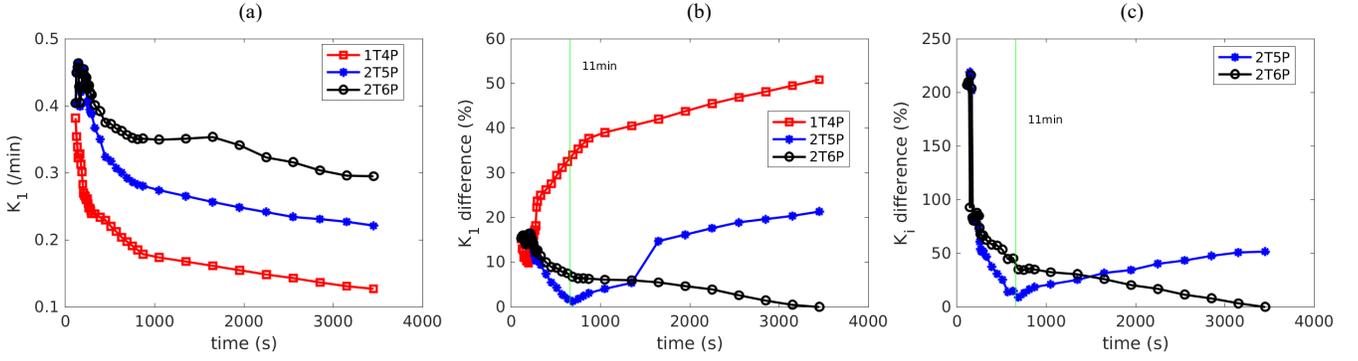

**Figure 4**. Effect of scan duration on myocardial kinetic estimation in different kinetic models. **(a)** Plots of $K_1$ as a function of scan duration for different kinetic models for one example patient; **(b)** Mean absolute difference in $K_1$ quantification as compared to the reference $K_1$ values. The mean difference was calculated over 10 patients. **(c)** Mean absolute difference in $K_i$ quantification as compared to the reference $K_i$ values.

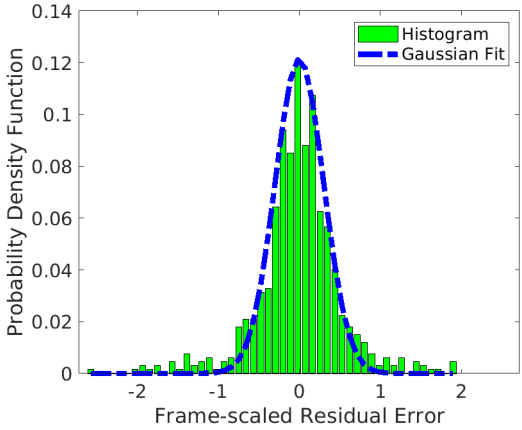

**Figure 5**. Histogram of normalized residual activity $\Delta C_m$ from the dynamic FDG-PET scans of 14 patients and a fit with the Gaussian distribution.

The plots in Figure 4(a) and 4(b) also show that the use of the 2T5P model for the one-hour data resulted in a more than 20% underestimation in $K_1$ as compared to the reference value, indicating that neglecting $k_4$ can actually affect the estimation of $K_1$. However, at short scan durations the error in $K_1$ by the 2T5P model decreases as the scan duration increases and reaches its minimum 2% at around 10-15 minutes.

The $K_1$ estimated by 1T4P significantly deviated from the reference $K_1$ when the duration was longer than 3-4 minutes, mainly due to the neglect of the phosphorylation parameters ($k_3$ and/or $k_4$) in the 1T4P. The mean absolute difference in $K_1$ from 1T4P decreases as the scan duration increases and reaches its minimum 10% around 2-3 minutes.

Figure 4(c) shows the mean absolute difference for $K_i$ quantification by the different models. Again, the full one-hour data with the 2T6P model was used to provide the reference $K_i$ value in each patient. Note that the 1T4P model cannot be used to quantify $K_i$. Overall, the effect of scan duration on $K_i$ quantification was very similar to that of $K_1$. But the error from the 2T5P model was larger, over 50% for the full one-hour data and 8% for the shortened scan of 10-15 minutes.

*F. Identifiability Analysis for FDG $K_1$ Quantification*

Figure 5 shows the histogram of the normalized residual activities from the dynamic FDG-PET scans of 10 patients. The histogram was well-matched with the zero-mean Gaussian with a standard deviation of $S_c = 0.3$. Hence, we used $S_c = 0.3$ to simulate noisy TACs in the practical identifiability analysis unless specified otherwise.

Figure 6 shows the plots of the bias and SD of $K_1$ as a function of scan duration for different models. The results here were averaged across multiple patients. Overall, the SD of $K_1$ estimated by different models generally followed the same trend – decreasing scan duration increases the SD. The minimum SD of 2T5P was lower than that of 2T6P (10% vs. 12%) because a smaller number of unknown parameters were estimated in the former model, which is associated with lower uncertainty. The 1T4P model led to a further reduction in the SD as compared to 2T5P and 2T6P.

The bias of $K_1$, however, behaved very differently in the three different models. The bias of $K_1$ by 2T6P gradually increased from 3% to 9% as the scan duration decreased from one hour to 2 minutes. In comparison, the bias of $K_1$ by 2T5P was 20% at one hour but decreased as the scan duration was shortened to 10 minutes where the bias reached its minimum of 3%. The bias then increased as the scan duration continued to decrease. The bias of $K_1$ by the 1T4P model was nearly 50% with the one-hour scan duration and gradually decreased to 6% as the scan duration was shortened to 2 minutes.

The changing trend on the bias of $K_1$ from this simulation study is consistent with the patient data presented in figure 3 and figure 4.

*G. Identifiability Analysis for Other Kinetic Parameters*

Figure 7 shows the mean absolute bias and SD of FDG $K_i$ for the 2T6P and 2T5P models. The 1T4P model cannot be used to provide $K_i$ estimation. The trend of the change in $K_i$ was very similar to that of $K_1$ in each of the two kinetic models. The minimum bias of $K_i$ was 4% by the 2T6P model but about 50% by the 2T5P model, indicating that the 2T5P model might be inappropriate for accurate $K_i$ quantification in the myocardium. The 2T6P and full one-hour data are instead required.

Table 2 summarizes the mean absolute bias, SD, and RMSE of multiple kinetic parameters estimated from the 2-minute

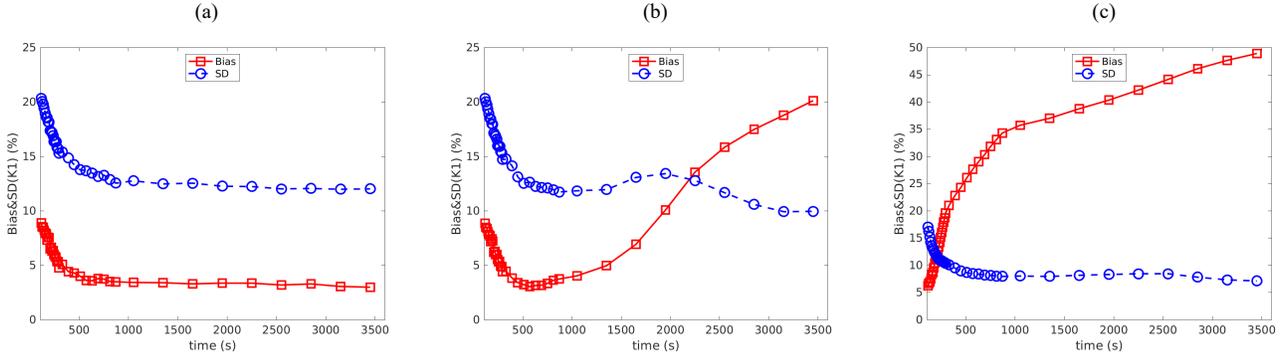

Figure 6. Plots of the mean absolute bias and SD of $K_1$ as a function of scan duration for three different kinetic models: (a) 2T6P, (b) 2T5P, and (c) 1T4P. The 1T4P model cannot be used to estimate a $K_i$. The scan duration is varied from 2 minutes to one hour. The mean bias and SD were calculated from ten applicable patients.

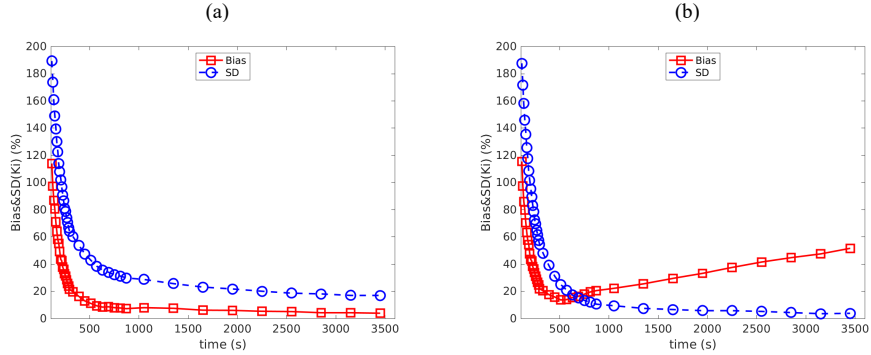

Figure 7. Plots of the mean absolute bias and SD of $K_i$ as a function of scan duration for two different kinetic models: (a) 2T6P and (b) 2T5P. The scan duration is varied from 2 minutes to one hour.

(2M) early dynamic scan with the 1T4P model, the 10-minute (10M) early-dynamic data with the 2T5P model, and the 1-hour (1H) dynamic data with the 2T6P model. The three protocols represent the appropriately optimal option for a short, medium, and long scan duration. For $K_1$ quantification, the 2T5P-10M protocol achieved similar bias, SD and RMSE as compared to the full 2T6P-1H protocol. While the 1T4P-2M protocol had a 2 times higher bias, the bias still remained below 10%. For $K_i$ quantification, the 2T5P protocol did not show comparable performance. The bias was nearly 3 times the bias of the 2T6P-1H protocol for $K_i$.

Table 2 also shows the results for other parameters including $k_2$, $k_3$, and $k_4$. The bias and SD of $k_3$ are generally higher than that of $K_1$ and $K_i$, indicating it is less stable for clinical use.

### H. Effect of Noise Level

Fig. 8 shows the patient-averaged bias and SD of FDG $K_1$ as a function of the noise level $S_c$ for the 2T6P-1H, 2T5P-10M, and 1T4P-2M protocols. For the GE Discovery ST scanner that we used in this patient study, the noise level in the global myocardium corresponds to $S_c = 0.3$ and the noise level in the myocardial segments of the AHA 17-segment model appropriately corresponds to $S_c = 1.0$. For the 2T6P-1H, as the noise level increases, both the bias and SD increase in $K_1$ quantification. At a low noise level ($S_c = 0.3$), $K_1$ had a bias of 3%. At the high noise level $S_c = 1.0$, the bias of $K_1$ became over 15%. The $K_1$ derived by 2T5P-10M had a slightly better bias curve than the 2T6P-1H. Because of model mismatch, the 1T4P-2M protocol had higher bias for all noise levels.

Table 2. Mean absolute bias, SD, and RMSE of different kinetic parameters estimated by different kinetic models in all patients. 2M, 10M, and 1H refer to 2 minutes, 10 minutes, and 1 hour, respectively.

| KINETIC PARAMETER | MODEL & TIME | BIAS (%) | SD (%) | RMSE (%) |
|---|---|---|---|---|
| $K_1$ | 1T4P, 2 M | 6.2 | 17.0 | 18.9 |
|  | 2T5P, 10 M | 3.1 | 12.6 | 13.2 |
|  | 2T6P, 1 H | 3.0 | 12.0 | 12.6 |
| $K_i$ | 1T4P, 2 M | / | / | / |
|  | 2T5P, 10 M | 13.9 | 20.9 | 28.1 |
|  | 2T6P, 1 H | 3.9 | 17.0 | 17.6 |
| $k_2$ | 1T4P, 2M | 12.5 | 29.0 | 34.3 |
|  | 2T5P, 10M | 5.0 | 15.7 | 16.9 |
|  | 2T6P, 1H | 4.0 | 14.6 | 15.4 |
| $k_3$ | 1T4P, 2M | / | / | / |
|  | 2T5P, 10M | 17.2 | 28.8 | 36.2 |
|  | 2T6P, 1H | 6.5 | 27.2 | 28.2 |
| $k_4$ | 1T4P, 2M | / | / | / |
|  | 2T5P, 10M | / | / | / |
|  | 2T6P, 1H | 3.4 | 21.5 | 22.0 |

The SD of $K_1$ quantification with 2T6P-1H was 12% at the low noise level of $S_c = 0.3$ and 40% at the high noise level of $S_c = 1.0$. The 2T5P-10M protocol achieved a very similar noise performance as the 2T6P-1H protocol, though the 1T4P-

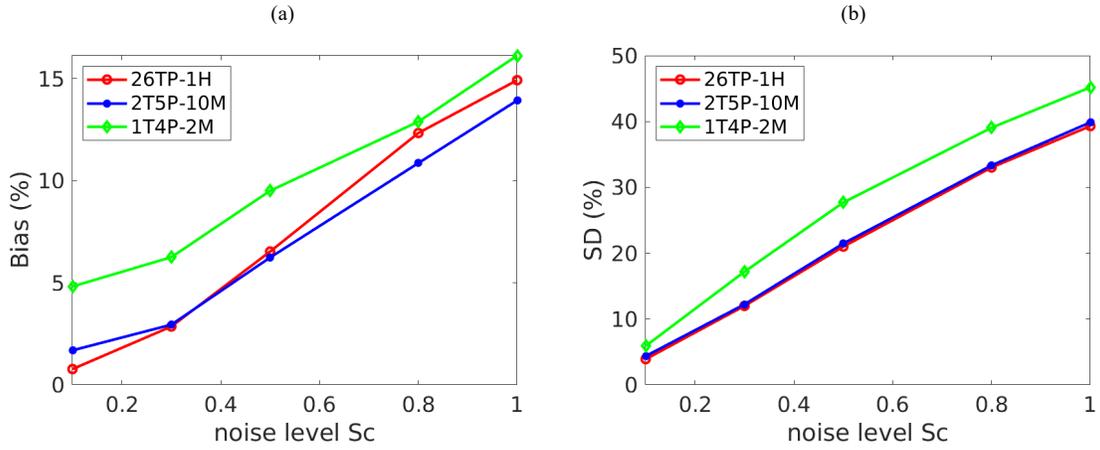

**Figure 8**. Plots of patient-averaged bias and standard deviation (SD) of myocardial FDG $K_1$ as a function of noise level $S_c$ for the 2T6P-1H, 2T5P-10M, and 1T4P-2M protocols. **(a)** Bias in percentage; **(b)** SD in percentage.

2M protocol was associated with higher SD at all noise levels. The bias and SD results indicate the performance of 2T5P-10M is comparable to that of 2T6P-1H for FDG $K_1$ quantification, while the 1T4P-2M protocol is less similar.

## IV. DISCUSSION

In this paper, we investigated tracer kinetic modeling strategies for dynamic cardiac FDG-PET for kinetic quantification of myocardial kinetics with emphasis on glucose transport. Both the reversible and irreversible two-tissue models were used in previous studies [22, 34]. However, the majority of existing dynamic cardiac FDG-PET focused on glucose metabolism ($K_i$) evaluation [16-18, 22, 27, 43], few on glucose transport ($K_1$) evaluation. The results from our study indicate that the reversible model (2T6P) is preferred for accurate modeling of one-hour dynamic cardiac FDG-PET data in humans (Fig. 3 and Fig. 4), suggesting that the dephosphorylation process (i.e., nonzero $k_4$) is non-negligible in kinetic modeling of one-hour dynamic cardiac FDG-PET data. The bias and SD associated with 2T6P of one-hour data were about 3% and 12% for global myocardial $K_1$ quantification, and 4% and 17% for global $K_i$ quantification, respecitvely (Table 2).

When analyzing the one-hour dynamic data, the use of the irreversible model (2T5P) could result in a large bias in FDG $K_i$ (up to 50%) for glucose metabolism evaluation (Fig. 7b) and in FDG $K_1$ (up to 20%) for glucose transport evaluation (Fig. 6b). The $K_i$ results from 2T5P and 2T6P in figure 7 are consistent with general consensus that a more complex model is more accurate (lower bias) but can be less stable (higher SD), see [44] for an example in cardiac imaging.

However, the irreversible model can become increasing accurate for FDG $K_1$ quantification if the scan duration is reduced from one hour to about 10-15 minutes (Fig. 6b). The estimation of the dephosphorylation rate $k_4$ becomes negligible for the shortened scan duration. The associated bias and SD with $K_1$ quantification was 3% and 13% (Table 2), respectively, which are comparable to the accuracy achieved by the 2T6P of the one-hour data. This suggests it is possible to quantify $K_1$ accurately without using a one-hour scan duration but a much-shortened scan with an appropriate kinetic model.

When the scan duration is further reduced to 2-3 minutes, the 1T4P model can become appropriate (Fig. 3(c) and Fig. 6(c)), though is associated with a non-negligible bias in FDG $K_1$ (~6%) (Fig. 6c and Table 2). These results on 2T6P, 2T5P, and 1T4P confirmed that the kinetic model selection depends on the scan duration of dynamic PET imaging.

The finding on the $K_1$ quantification accuracy of the 2T5P model along with a scan duration of about 10-15 minutes is significant and can guide our future protocol designs for testing and applying single-tracer multiparametric imaging in clinics. FDG $K_1$ has the potential to be used as a potential surrogate of myocardial blood flow for simultaneous flow-metabolism evaluation or is used directly for evaluating a transport-metabolism relationship. There are potentially two implementation options. The first option is to use the full one-hour dynamic FDG scanning, which can provide both $K_1$ and $K_i$ using the 2T6P model. A potential benefit of this protocol is that quantitative $K_i$ is potentially more beneficial than the semi-quantitative SUV for the characterization of myocardial glucose metabolism. The disadvantage is that compared to standard static scanning, the prolonged dynamic scan duration not only increases the scan cost but also may result in compromised dynamic image quality or image degradation due to patient movement during the long scan.

The other option is to add an early-dynamic scan of about 10-15 minutes right after the FDG injection. This add-on scan protocol is only used for FDG $K_1$ quantification. Evaluation of myocardial glucose metabolism is still achieved using the standard clinical protocol, i.e., static scanning to provide a late-time SUV usually at 75-95 minutes post-injection. One potential advantage of this protocol is it would only occupy an additional scanner time of 10-15 minutes and allow the same scanner to be used for other patients in between the two scans.

This study has several limitations. While our ultimate goal is to develop a single tracer multiparametric imaging method using tracer kinetic modeling, this work only focused on the first technical step to identify the appropriate method and protocol for kinetic quantification without directly evaluating the impact for clinical assessment. The limitations also include that we did not have a ground truth for the kinetic parameter quantification, although this is a general challenge for any study using patient data. We instead used the estimates of the 2T6P model from the full one-hour dynamic data as the reference. To validate the findings of the patient study, we had further conducted a computer simulation study to analyze the identifiability of the kinetic model parameters.

The analysis of this study was limited to evaluation of global myocardial quantification, not investigating quantification at the level of myocardial segments or parametric imaging. In the disease cohort, a global myocardial TAC can be a mix of normal and abnormal tissue signals, thus neglecting the potential wide physiological heterogeneity. The major reason that we did not include segment-level investigation was the dynamic data of segments in this study are very noisy. The scanner used in this study is relatively old (2002 GE Discovery ST model) with no time-of-flight (TOF) capability and as a result, noise performance was far from optimal for exploring quantification in myocardial segments. The result from the practical identifiability analysis indicates it is less reliable to do FDG $K_1$ quantification in individual myocardial segments as the resulting bias and SD are high (Fig. 8; corresponding to $S_c = 1.0$).

The study did not consider motion correction, which could affect the accuracy of kinetic quantification [45-48]. Nevertheless, the effect of motion is not expected to result in a significant change to the results of this study since the spatial resolution of the PET scanner is only about 6-8 mm and the quantification was performed on large, global ROIs.

There are a number of options to address the limitations of the study in our future work. Compared to the GE Discovery ST scanner used in this study, latest scanners (e.g., Siemens Biograph Vision [49], GE Discovery MI [50]) have much higher scanner sensitivity which, together with time-of-flight capability, may reduce the noise at the myocardial segmental level from $S_c = 1.0$ to $S_c = 0.3$ or lower to allow FDG kinetic quantification in individual segments. The improved spatial resolution can also justify the increasing need for motion correction. In particular, the UIH uEXPLORER total-body PET/CT scanner [51, 52] has further improved sensitivity that may make it feasible to implement parametric imaging [53] in the myocardium. In addition, the images were reconstructed using standard vendor reconstruction in this study. Improved image reconstruction has been developed for dynamic PET imaging [53, 54] with the kernel methods [55, 56] as one of recent examples. Thus, our future work will explore the use of the latest PET/CT scanners and/or advanced image reconstruction to improve data quality for segment-level and voxel-wise kinetic quantification in the myocardium.

## V. Conclusions

In tracer kinetic modeling of dynamic cardiac FDG-PET, optimal kinetic model selection depends on scan duration. The reversible two-tissue model, irreversible two-tissue model, and simplified one-tissue model are respectively appropriate for analyzing dynamic FDG imaging with a scan duration of about one-hour, 10-15 minutes, and less than 2-3 minutes. An early dynamic scan of about 10-15 minutes with irreversible kinetic modeling can be comparable to the full one-hour scan with reversible kinetic modeling for the evaluation of glucose transport rate $K_1$ in the myocardium. Our future work will further improve the technical method and compare FDG $K_1$ to myocardial blood flow using patient data.

## VI. Acknowledgments

The authors thank Denise Caudle, Michael Rusnak, and Ben Spencer for their assistance in the dynamic PET/CT data acquisition, Diana Ramos for her efforts in patient recruitments, and the patients that agreed to participate in these studies.

## References

[1] S. R. Cherry, J. A. Sorenson, and M. E. Phelps, *Physics in Nuclear Medicine E-Book*. Elsevier Health Sciences, 2012.

[2] J. W. Fletcher *et al.*, "Recommendations on the use of F-18-FDG PET in oncology," *Journal of Nuclear Medicine,* vol. 49, no. 3, pp. 480-508, Mar 2008, doi: 10.2967/jnumed.107.047787.

[3] H. Schelbert, M. Phelps, C. Selin, E. Hoffman, and D. Kuhl, "Glucose-Metabolism of Regional Myocardial Ischemia Evaluated by 18 Fluoro-2-Deoxyglucose and Positron Emission Computed-Tomography," (in English), *American Journal of Cardiology,* vol. 45, no. 2, pp. 465-465, 1980, doi: Doi 10.1016/0002-9149(80)90967-4.

[4] P. G. Camici, S. K. Prasad, and O. E. Rimoldi, "Stunning, hibernation, and assessment of myocardial viability," *Circulation,* vol. 117, no. 1, pp. 103-114, Jan 2008, doi: 10.1161/circulationaha.107.702993.

[5] G. T. Gullberg, U. M. Shrestha, and Y. Seo, "Dynamic cardiac PET imaging: Technological improvements advancing future cardiac health," *Journal of Nuclear Cardiology,* vol. 26, no. 4, pp. 1292-1297, Aug 2019, doi: 10.1007/s12350-018-1201-3.

[6] G. El Fakhri *et al.*, "Reproducibility and Accuracy of Quantitative Myocardial Blood Flow Assessment with (82)Rb PET: Comparison with (13)N-Ammonia PET," *Journal of Nuclear Medicine,* vol. 50, no. 7, pp. 1062-1071, Jul 2009, doi: 10.2967/jnumed.104.007831.

[7] S. V. Nesterov *et al.*, "Quantification of Myocardial Blood Flow in Absolute Terms Using Rb-82 PET Imaging The RUBY-10 Study," *Jacc-Cardiovascular Imaging,* vol. 7, no. 11, pp. 1119-1127, Nov 2014, doi: 10.1016/j.jcmg.2014.08.003.

[8] J. M. Renaud, J. N. DaSilva, R. S. B. Beanlands, and R. A. deKemp, "Characterizing the normal range of myocardial blood flow with (82)rubidium and N-13-ammonia PET imaging," *Journal of Nuclear Cardiology,* vol. 20, no. 4, pp. 578-591, Aug 2013, doi: 10.1007/s12350-013-9721-3.

[9] J. B. Moody, B. C. Lee, J. R. Corbett, E. P. Ficaro, and V. L. Murthy, "Precision and accuracy of clinical quantification of myocardial blood flow by dynamic PET: A technical perspective," *Journal of Nuclear Cardiology,* vol. 22, no. 5, pp. 935-951, Oct 2015, doi: 10.1007/s12350-015-0100-0.

[10] V. L. Murthy *et al.*, "Comparison and Prognostic Validation of Multiple Methods of Quantification of Myocardial Blood Flow with Rb-82 PET," *Journal of Nuclear Medicine,* vol. 55, no. 12, pp. 1952-1958, Dec 2014, doi: 10.2967/jnumed.114.145342.

[11] H. Iida *et al.*, "Noninvasive quantitation of cerebral blood flow using oxygen-15-water and a Dual-PET system," *Journal of Nuclear Medicine,* vol. 39, no. 10, pp. 1789-1798, Oct 1998. [Online]. Available: <Go to ISI>://WOS:000076316200032.

[12] H. J. Harms *et al.*, "Comparison of clinical non-commercial tools for automated quantification of myocardial blood flow using oxygen-15-labelled water PET/CT," *European Heart Journal-Cardiovascular Imaging,* vol. 15, no. 4, pp. 431-441, Apr 2014, doi: 10.1093/ehjci/jet177.

[13] T. C. Rust, E. V. R. DiBella, C. J. McGann, P. E. Christian, J. M. Hoffman, and D. J. Kadrmas, "Rapid dual-injection single-scan N-13-ammonia PET for quantification of rest and stress myocardial blood flows," *Physics in Medicine and Biology,* vol. 51, no. 20, pp. 5347-5362, Oct 2006, doi: 10.1088/0031-9155/51/20/018.

[14] A. Abraham *et al.*, "F-18-FDG PET Imaging of Myocardial Viability in an Experienced Center with Access to F-18-FDG and Integration with Clinical Management Teams: The Ottawa-FIVE Substudy of the PARR 2 Trial," *Journal of Nuclear Medicine,* vol. 51, no. 4, pp. 567-574, Apr 2010, doi: 10.2967/jnumed.109.065938.

[15] P. G. Camici and O. E. Rimoldi, "The clinical value of myocardial blood flow measurement," *J Nucl Med,* vol. 50, no. 7, pp. 1076-87, Jul 2009, doi: 10.2967/jnumed.108.054478.

[16] A. P. van der Weerdt, L. J. Klein, R. Boellaard, C. A. Visser, F. C. Visser, and A. A. Lammertsma, "Image-derived input functions for determination of MRGlu in cardiac F-18-FDG PET scans," *Journal of Nuclear Medicine,* vol. 42, no. 11, pp. 1622-1629, Nov 2001. [Online]. Available: <Go to ISI>://WOS:000172090900016.

[17] S. Y. Tsai *et al.*, "Clinical significance of quantitative assessment of right ventricular glucose metabolism in patients with heart failure with reduced ejection fraction," *European Journal of Nuclear Medicine and Molecular Imaging,* vol. 46, no. 12, pp. 2601-2609, Nov 2019, doi: 10.1007/s00259-019-04471-9.

[18] K. Y. Ko *et al.*, "Clinical significance of quantitative assessment of glucose utilization in patients with ischemic cardiomyopathy," *Journal of Nuclear Cardiology,* vol. 27, no. 1, pp. 269-279, Feb 2020, doi: 10.1007/s12350-018-1395-4.

[19] A. Bertoldo, P. Peltoniemi, V. Oikonen, J. Knuuti, P. Nuutila, and C. Cobelli, "Kinetic modeling of F-18 FDG in skeletal muscle by PET: a four-compartment five-rate-constant model," *American Journal of Physiology-Endocrinology and Metabolism,* vol. 281, no. 3, pp. E524-E536, Sep 2001. [Online]. Available: <Go to ISI>://WOS:000170569900014.

[20] D. Annane *et al.*, "Correlation between decreased myocardial glucose phosphorylation and DNA mutation size in myotonic dystrophy," *Circulation,* vol. 90, no. 6, pp. 2629-2634, Dec 1994. [Online]. Available: <Go to ISI>://WOS:A1994PX37400005.

[21] X. Li, D. Feng, K. P. Lin, and S. C. Huang, "Estimation of myocardial glucose utilisation with PET using the left ventricular time-activity curve as a non-invasive input function," *Medical & Biological Engineering & Computing,* vol. 36, no. 1, pp. 112-117, Jan 1998, doi: 10.1007/bf02522867.

[22] K. Morita *et al.*, "Quantitative analysis of myocardial glucose utilization in patients with left ventricular dysfunction by means of F-18-FDG dynamic positron tomography and three-compartment analysis," *European Journal of Nuclear Medicine and Molecular Imaging,* vol. 32, no. 7, pp. 806-812, Jul 2005, doi: 10.1007/s00259-004-1743-2.

[23] Y. Choi *et al.*, "Parametric images of myocardial metabolic rate of glucose generated from dynamic cardiac PET and 2-[18F]fluoro-2-deoxy-d-glucose studies," *Journal of Nuclear Medicine,* vol. 32, no. 4, pp. 733-738, Apr 1991. [Online]. Available: <Go to ISI>://WOS:A1991FG79700031.

[24] M. Taylor *et al.*, "An evaluation of myocardial fatty acid and glucose uptake using PET with F-18 fluoro-6-thia-heptadecanoic acid and F-18 FDG in patients with congestive heart failure," *Journal of Nuclear Medicine,* vol. 42, no. 1, pp. 55-62, Jan 2001. [Online]. Available: <Go to ISI>://WOS:000166429400031.

[25] P. Herrero, T. L. Sharp, C. Dence, B. M. Haraden, and R. J. Gropler, "Comparison of 1-C-11-glucose and F-18-FDG for quantifying myocardial glucose use with PET," *Journal of Nuclear Medicine,* vol. 43, no. 11, pp. 1530-1541, Nov 2002. [Online]. Available: <Go to ISI>://WOS:000179112100019.

[26] H. Wiggers, M. Bottcher, T. T. Nielsen, A. Gjedde, and H. E. Botker, "Measurement of myocardial glucose uptake in patients with ischemic cardiomyopathy: Application of a new quantitative method using regional tracer kinetic information," *Journal of Nuclear Medicine,* vol. 40, no. 8, pp. 1292-1300, Aug 1999. [Online]. Available: <Go to ISI>://WOS:000081902300011.

[27] A. Lebasnier *et al.*, "Diagnostic value of quantitative assessment of cardiac F-18-fluoro-2-deoxyglucose uptake in suspected cardiac sarcoidosis," *Annals of Nuclear Medicine,* vol. 32, no. 5, pp. 319-327, Jun 2018, doi: 10.1007/s12149-018-1250-3.

[28] A. Cochet *et al.*, "Evaluation of Breast Tumor Blood Flow with Dynamic First-Pass F-18-FDG PET/CT: Comparison with Angiogenesis Markers and Prognostic Factors," *Journal of Nuclear Medicine,* vol. 53, no. 4, pp. 512-520, Apr 2012, doi: 10.2967/jnumed.111.096834.

[29] N. A. Mullani, R. S. Herbst, R. G. O'Neil, K. L. Gould, B. J. Barron, and J. L. Abbruzzese, "Tumor blood flow measured by PET dynamic imaging of first-pass F-18-FDG uptake: A comparison with O-15-Labeled water-measured blood flow," *J. Nucl. Med.,* vol. 49, no. 4, pp. 517-523, Apr 2008, doi: 10.2967/jnumed.107.048504.

[30] H. Bernstine *et al.*, "FDG PET/CT Early Dynamic Blood Flow and Late Standardized Uptake Value Determination in Hepatocellular Carcinoma," *Radiology,* vol. 260, no. 2, pp. 503-510, Aug 2011, doi: 10.1148/radiol.11102350.

[31] J. Maddahi, "Properties of an ideal PET perfusion tracer: New PET tracer cases and data," *Journal of Nuclear Cardiology,* vol. 19, pp. S30-S37, Feb 2012, doi: 10.1007/s12350-011-9491-8.

[32] M. F. Di Carli, S. Dorbala, J. Meserve, G. El Fakhri, A. Sitek, and S. C. Moore, "Clinical myocardial perfusion PET/CT," (in English), *Journal of Nuclear Medicine,* vol. 48, no. 5, pp. 783-793, May 2007, doi: 10.2967/jnumed.106.032789.

[33] L. K. Shankar *et al.*, "Consensus recommendations for the use of F-18-FDG PET as an indicator of therapeutic response in patients in national cancer institute trials," *Journal of Nuclear Medicine,* vol. 47, no. 6, pp. 1059-1066, Jun 2006. [Online]. Available: <Go to ISI>://WOS:000238104800028.

[34] M. E. Phelps, S. C. Huang, E. J. Hoffman, C. Selin, L. Sokoloff, and D. E. Kuhl, "Tomographic measurement of local cerebral glucose metabolic rate in humans with (F-18)2-fluoro-2-deoxy-D-glucose: validation of method," *Ann Neurol,* vol. 6, no. 5, pp. 371-88, Nov 1979, doi: 10.1002/ana.410060502.

[35] R. E. Carson, "Tracer Kinetic Modeling in PET," in *Positron Emission Tomography*, D. L. Bailey, D. W. Townsend, P. E. Valk, and M. N. Maisey Eds. London: Springer, 2005.

[36] E. D. Morris, C. J. Endres, K. C. Schmidt, B. T. Christian, R. F. Muzic, and R. E. Fisher, "Kinetic modeling in positron emission tomography," in *Emission Tomography: The Fundamentals of PET and SPECT*, Wermick MN and A. JN Eds.: Elsevier Inc., 2004, pp. 499-540.

[37] G. Wang and J. Qi, "An Optimization Transfer Algorithm for Nonlinear Parametric Image Reconstruction From Dynamic PET Data," *IEEE Transactions on Medical Imaging,* vol. 31, no. 10, pp. 1977-1988, Oct 2012, doi: 10.1109/tmi.2012.2212203.

[38] G. Glatting, P. Kletting, S. N. Reske, K. Hohl, and C. Ring, "Choosing the optimal fit function: Comparison of the Akaike information criterion and the F-test," *Medical Physics,* vol. 34, no. 11, pp. 4285-4292, Nov 2007, doi: 10.1118/1.2794176.

[39] M. A. Richard, J. P. Fouquet, R. Lebel, and M. Lepage, "Determination of an Optimal Pharmacokinetic Model of F-18-FET for Quantitative Applications in Rat Brain Tumors," *Journal of Nuclear Medicine,* vol. 58, no. 8, pp. 1278-1284, Aug 2017, doi: 10.2967/jnumed.116.180612.

[40] K. P. Burnham and D. R. Anderson, "Multimodel inference - understanding AIC and BIC in model selection," *Sociological Methods & Research,* vol. 33, no. 2, pp. 261-304, Nov 2004, doi: 10.1177/0049124104268644.

[41] Y. Zuo, S. Sarkar, M. T. Corwin, K. Olson, R. D. Badawi, and G. B. Wang, "Structural and practical identifiability of dual-input kinetic modeling in dynamic PET of liver inflammation," *Physics in Medicine and Biology,* vol. 64, no. 17, Sep 2019, Art no. 175023, doi: 10.1088/1361-6560/ab1f29.

[42] Y. J. Wu and R. E. Carson, "Noise reduction in the simplified reference tissue model for neuroreceptor functional Imaging," *Journal of Cerebral Blood Flow and Metabolism,* vol. 22, no. 12, pp. 1440-1452, Dec 2002, doi: 10.1097/01.wcb.0000033967.83623.34.

[43] Y. W. Wu, S. Y. Wang, K. M. Chiu, and S. H. Chu, "Quantitative analysis of myocardial glucose metabolism by using dynamic FDG PET in patients with ischemic cardiomyopathy," *Journal of Nuclear Medicine,* vol. 55, May 2014. [Online]. Available: <Go to ISI>://WOS:000361438102448.

[44] A. M. Alessio, J. B. Bassingthwaighte, R. Glenny, and J. H. Caldwell, "Validation of an axially distributed model for quantification of myocardial blood flow using $^{13}$N-ammonia PET," *J Nucl Cardiol,* vol. 20, no. 1, pp. 64-75, Feb 2013, doi: 10.1007/s12350-012-9632-8.

[45] Y. Petibon, T. Sun, P. K. Han, C. Ma, G. El Fakhri, and J. Ouyang, "MR-based cardiac and respiratory motion correction of PET: application to static and dynamic cardiac F-18-FDG imaging," *Phys. Med. Biol.,* vol. 64, no. 19, Oct 2019, Art no. 195009, doi: 10.1088/1361-6560/ab39c2.

[46] A. Poitrasson-Riviere *et al.*, "Reducing motion-correction-induced variability in (82)rubidium myocardial blood-flow quantification," *Journal of Nuclear Cardiology*, doi: 10.1007/s12350-019-01911-9.

[47] C. Byrne, A. Kjaer, N. E. Olsen, J. L. Forman, and P. Hasbak, "Test-retest repeatability and software reproducibility of myocardial flow measurements using rest/adenosine stress Rubidium-82 PET/CT with and without motion correction in healthy young volunteers," *Journal of Nuclear Cardiology*, doi: 10.1007/s12350-020-02140-1.

[48] Y. H. Lu *et al.*, "Data-driven voluntary body motion detection and non-rigid event-by-event correction for static and dynamic PET," *Phys. Med. Biol.,* vol. 64, no. 6, Mar 2019, Art no. 065002, doi: 10.1088/1361-6560/ab02c2.

[49] J. van Sluis *et al.*, "Performance characteristics of the digital Biograph Vision PET/CT system," *Journal of Nuclear Medicine,* vol. 60, no. 7, pp. 1031-1036, 2019.

[50] T. Pan *et al.*, "Performance evaluation of the 5-Ring GE Discovery MI PET/CT system using the national electrical manufacturers association NU 2-2012 Standard," *Medical Physics,* vol. 46, no. 7, pp. 3025-3033, 2019.

[51] S. R. Cherry, T. Jones, J. S. Karp, J. Y. Qi, W. W. Moses, and R. D. Badawi, "Total-Body PET: Maximizing Sensitivity to Create New Opportunities for Clinical Research and Patient Care," *Journal of Nuclear Medicine,* vol. 59, no. 1, pp. 3-12, Jan 2018, doi: 10.2967/jnumed.116.184028.

[52] R. Badawi *et al.*, "First Human Imaging Studies with the EXPLORER Total-Body PET Scanner," *Journal of Nuclear Medicine,* vol. 60, no. 3, pp. 299-303, 2019.

[53] J. D. Gallezot, Y. H. Lu, M. Naganawa, and R. E. Carson, "Parametric Imaging With PET and SPECT," *IEEE Transactions on Radiation and Plasma Medical Sciences,* vol. 4, no. 1, pp. 1-23, 2020.

[54] A. J. Reader and J. Verhaeghe, "4D image reconstruction for emission tomography," *Physics in Medicine and Biology,* vol. 59, no. 22, pp. R371-R418, Nov 21 2014, doi: 10.1088/0031-9155/59/22/r371.

[55] G. Wang and J. Qi, "PET image reconstruction using kernel method," *IEEE Transactions on Medical Imaging,* vol. 34, no. 1, pp. 61-71, 2015.

[56] G. B. Wang, "High temporal-resolution dynamic PET image reconstruction using a new spatiotemporal kernel method," *IEEE Transactions on Medical Imaging,* vol. 38, no. 3, pp. 664 – 674, 2019.